# Molecular Interactions in Chiral Nematic Liquid Crystals and Enantiotopic Discrimination through the NMR Spectra of Achiral Molecules I: Rigid Solutes

*Anant Kumar, Alexandros G. Vanakaras[1*] and Demetri J. Photinos*

*Department of Materials Science, University of Patras, Patras 26504, Greece*

**ABSTRACT:**

We have developed a molecular theory for the enantiotopic discrimination in prochiral solutes dissolved in chiral nematic solvents by means of NMR spectroscopy. The leading rank tensor contributions to the proposed potential of mean torque include symmetric as well as antisymmetric terms with respect to spatial inversion, these lead to a consistent determination of all the prochiral solute symmetries for which enantiotopes are distinguishable by NMR and also to excellent quantitative agreement when tested against the available experimental data for the rigid solute acenaphthene and for the moderately flexible ethanol.

1. **INTRODUCTION**

Nuclear Magnetic Resonance (NMR) spectroscopy of molecules dissolved in liquid crystals (LCs) can provide, through the so called segmental order parameters, quantitative information on the orientational ordering of various segments of the solute molecules and thereby to elucidate their structure and conformations [1]. Conversely, the use of solutes of known structure can often provide information on the molecular organization of the solvent phase. Liquid crystal solvents, unlike isotropic liquids, prevent the anisotropic magnetic interactions from averaging out completely, thus allowing their spectral evaluation through nuclear dipole-dipole coupling, nuclear quadrupole-electric field gradient coupling, chemical shift anisotropy etc. As NMR spectroscopy is insensitive to positional ordering, the majority of LC-NMR studies, particularly those aiming at determining the structure of solute molecules, use nematic LCs as solvents. Achiral nematics (N), being the structurally simplest of all liquid crystals, have been extensively used as solvents. They are however, "chirality blind" since a chiral solute and its enantiomer yield identical NMR spectra in such achiral solvents. In chiral nematic (N*) solvents, on the other hand, chiral enantiomeric pairs of solutes can exhibit measurably different spectra.

---

[1] Email: a.g.vanakaras@upatras.gr

An interesting situation arises when achiral solute molecules are dissolved in N* solvents. In such cases, if the achiral solute symmetry corresponds to an improper point group (i.e. containing reflection or roto-reflection symmetry elements) the NMR spectrum could discriminate between solute molecular fragments (single atoms, pairs of atoms, etc) which are related by improper symmetries while the part of the spectrum associated with the remaining molecular fragments would be qualitatively the same with the spectrum obtained in an achiral N solvent. Molecules of improper point group symmetry are called prochiral, as they would become chiral on replacing one of the improper symmetry related molecular elements by a different one. Molecular fragments which are related by improper symmetry operations are termed enantiotopic and the possible differentiation between such fragments through their NMR spectra is referred to as spectral enantiotopic discrimination. Such discrimination is also described as a lifting of the degeneracy in the NMR spectra of enantiotopes on replacing an achiral solvent with a chiral one.

The first demonstration that the NMR spectra associated with enantiotopic fragments of prochiral molecules dissolved in a N* solvent are not identical dates back to the late 60s [2]. The use of polypeptide N* and other lyotropic chiral nematic solvents was soon found to provide enhanced enantiotopic resolution [3], [4], [5]. Moreover, the spontaneously formed helical structure of the N* phase [6] was found to transform to a uniformly aligned structure under the action of the NMR spectrometer field, thus providing well resolved and readily analyzable spectra [7]. Since then, organic solutions of poly-γ-benzyl-L-glutamate (PBLG) have been the most commonly used chiral nematic solvent and the phenomenon of enantiotopic discrimination in chiral LCs has been attracting constant research interest; the numerous studies on the subject have recently been reviewed in a comprehensive article [8]. Enantiotropic discrimination of prochiral molecules has also been observed [9], [10] in the spontaneously twisted nematic phase ($N_X$) formed by achiral symmetric dimers [11], [12].The symmetry of this phase and the molecular organization therein differ from the chiral nematic (N*) phase composed of chiral molecules, to which we limit our attention in this work.

Despite the accumulated wealth of experimental NMR information on enantiotopic discrimination, its rationalization in terms of molecular interactions remains at a primitive stage. The presently used tools for the analysis and interpretation of the spectra obtained from prochiral solute molecules are based on the notion that their improper symmetry elements are "quenched" by the lower symmetry of the environment and thus the molecules acquire an "effective" symmetry which lacks some of the elements of the actual molecular symmetry [8], [13]. This change in the "effective" symmetry, when prochiral molecules are dissolved in chiral solvents, is then assumed to lead to an ordering tensor whose relative magnitudes of the principal values and the directions of the principal axes differ from those of the ordering tensor of the same molecule in an achiral solvent, wherein the actual molecular symmetry is understood to be retained.

In this work we start from the intermolecular interactions between a solvent-solute pair and, taking into account the symmetries of the chiral nematic solvent phase, we formulate the potential of mean torque that governs the orientational distribution function, and thereby the ordering, of a solute molecule. The leading contributions to the potential of mean torque are shown explicitly to be separable into inversion-symmetric terms, which are present in both chiral and achiral solvents, and inversion-antisymmetric terms which are present only in chiral solvents and with opposite signs for opposite solvent handedness. The possible symmetries of rigid prochiral molecules are considered in detail with regards to the chiral discrimination and its quantification in terms of solvent and solute interaction couplings, order parameters and geometry of the enantiotopic molecular segments. The proposed approach for the description of solute ordering in $N^*$ solvents is then applied to the reproduction of extensive experimental data which are available from dipolar and quadrupolar spectra of acenaphthene [14] and ethanol [15] molecules dissolved in chiral nematic solvents.

The formulation of the potential of mean torque is presented in Section 2 and the relation of the NMR spectra of the solute molecules to their order parameters in liquid crystalline solvents are given in Section 3. The existence of anantiotopic discrimination for different prochiral molecule symmetries is considered systematically in Section 4. The testing against experiment is presented in Section 5 and the conclusions drawn from this work are in Section 6.

## 2. POTENTIAL OF MEAN TORQUE IN A CHIRAL NEMATIC HOST

The potential of mean torque, $V(\omega)$, experienced by a rigid solute molecule dissolved in a positionally uniform and orientationally ordered liquid solvent is related the probability density $f(\omega)$ of finding the solute molecule in a given orientation $\omega$ according to [1].

$$f(\omega) = \zeta^{-1} \exp[-V(\omega)] \qquad . \qquad (2.1)$$

Here we formulate $V(\omega)$ for rigid achiral solutes of various symmetries and structures dissolved in chiral nematic ("cholesteric") liquid crystals [2]. The chiral nematic phase is understood to consist of chiral molecules, to have complete disorder in the positons of the constituent molecules and local orientational order of the $D_2$ point symmetry, implying three mutually orthogonal two-fold axes, identified with the phase directors $\hat{n}, \hat{l}, \hat{m}$. Therefore, strictly, the phase is locally biaxial. According to the $D_2$ symmetry, the local physical properties of the phase are invariant under simultaneous sign inversion of any two of the directors, as a result of its invariance under twofold rotations about the third director. With $\hat{n}$ identified as the primary nematic director, the biaxiality reflects the strict lack (however small) of equivalence of the other two directors, $\hat{l}$ and $\hat{m}$. In the helically twisted state of the chiral nematic liquid crystal, one of these two directors, say $\hat{l}$, is identified with the helix axis and the other, $\hat{m}$, remains on the plane perpendicular to the helix axis and follows the twisting of the primary director $\hat{n}$.

For the formulation of $V(\omega)$ we follow the usual approach of a tensor expansion in which a small number of leading rank terms are retained, guided by invariance and molecular symmetry requirements [1]. In particular, $V(\omega)$ is required to be a scalar function of the solute orientation $\omega$ and to remain invariant under the symmetry operations of the point groups of the solute and solvent molecules and of the solvent phase. To construct the possible tensor components of $V(\omega)$ which satisfy these requirements we define, in addition to the frame of solvent phase-fixed axes $\hat{n}, \hat{l}, \hat{m}$ a molecular axes frame $x, y, z$ for the solute molecules and an analogous frame $x', y', z'$ for the solvent molecules. The choice of axes for both these molecular frames is in principle arbitrary but the expression of $V(\omega)$ is considerably simplified if the directions of these axes are identified with specific symmetry elements of the molecules, when such elements are present. In our formulation we assume no particular symmetry for the (chiral) solvent molecules, while the solute molecules, due to their assumed achirality, will necessarily have certain symmetry elements, to be specified in Section 4. A solute and solvent pair is shown in Figure 1, with their molecular frames and the phase-fixed frame. In what follows, the molecular frame axes will be denoted generically by the unit vectors $\hat{a}, \hat{b} = \hat{x}, \hat{y}, \hat{z}$ and $\hat{a}', \hat{b}' = \hat{x}', \hat{y}', \hat{z}'$ and the phase directors by $\hat{I}, \hat{J} = \hat{n}, \hat{l}, \hat{m}$.

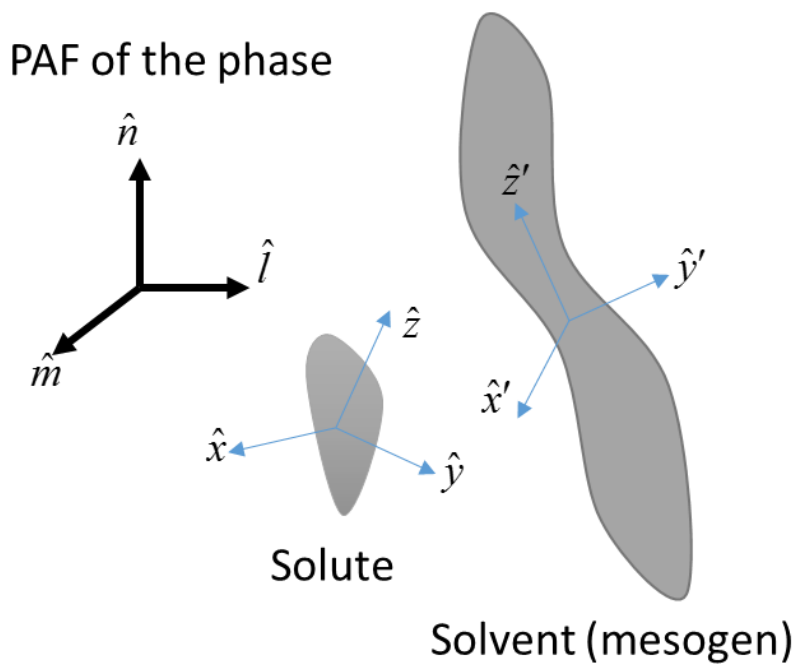

**Figure 1**: Orthogonal frames of axes of the solvent and solute molecules and the principal axes frame (PAF) of the solvent phase (director frame) used for the description of the potential of mean torque.

The lowest rank terms of $V(\omega)$ are of the "vector-vector" form

$$V^{(1)}(\omega) = \bar{C}^{(1)}_{a,a'} \langle \hat{a} \cdot \hat{a}' \rangle' \quad , \tag{2.2}$$

where $\bar{C}^{(1)}_{a,a'}$ are effective molecular coupling parameters, understood to be obtained after averaging the vector part of the solvent-solute interaction over the solvent molecule positions; the primed angular brackets indicate orientational averaging of the solvent molecule relative to the phase-fixed axes (director frame) and summation over repeated tensor indices/axis labels is implied throughout. Due to the D$_2$ symmetry of the phase, all the terms in eq(2.2) vanish. This becomes directly evident from $\langle \hat{a} \cdot \hat{a}' \rangle' = (\hat{a} \cdot \hat{I}) \langle \hat{I} \cdot \hat{a}' \rangle'$ and the vanishing of the solvent orientational averaging as a result of all three directors being axes of twofold symmetry of the phase. Thus there are no first rank (vector-vector) contributions to $V(\omega)$ for a chiral nematic (N*) solvent.

The symmetric second rank terms are of the form

$$V^{(2)}(\omega) = \bar{C}^{(2)}_{ab;a'b'} \left( \frac{3}{2} \langle (\hat{a} \cdot \hat{a}')(\hat{b} \cdot \hat{b}') \rangle' - \frac{1}{2} \delta_{ab} \delta_{a'b'} \right) \quad , \tag{2.3}$$

with the Kronecker delta $\delta_{ab} = 0, 1$ depending on whether $a \neq b$ or $a = b$ respectively. These are the routinely used starting terms for the formulation of $V(\omega)$ in the common achiral N phase [1]. Here, taking into account the D$_2$ symmetry of the N* phase, the complete second rank contribution of eq(2.3) can be put in the form

$$V^{(2)}(\omega) = \langle g^{(2)}_{ab} \rangle' S^n_{ab} + \langle \Delta^{(2)}_{ab} \rangle' \left( S^l_{ab} - S^m_{ab} \right) \quad , \tag{2.4}$$

where the orientational second rank tensor components $a,b$ for the director $\hat{I}$ are defined as

$$S^I_{ab} \equiv \frac{3}{2} (\hat{a} \cdot \hat{I})(\hat{b} \cdot \hat{I}) - \frac{1}{2} \delta_{ab} \; ; \; \hat{I} = \hat{n}, \hat{l}, \hat{m} \quad , \tag{2.5}$$

and the coupling coefficients to the solvent ordering $\langle g^{(2)}_{ab} \rangle'$ (uniaxial contribution) and $\langle \Delta^{(2)}_{ab} \rangle'$ (biaxial contribution) are defined in terms of the molecular coupling parameters $\bar{C}^{(2)}_{ab;a'b'}$ and the solvent order parameters $\langle S^I_{a'b'} \rangle'$ as follows:

$$\langle g^{(2)}_{ab} \rangle' \equiv \bar{C}^{(2)}_{ab;a'b'} \langle S^n_{a'b'} \rangle' \; ; \; \langle \Delta^{(2)}_{ab} \rangle' \equiv \frac{1}{3} \bar{C}^{(2)}_{ab;a'b'} \langle S^l_{a'b'} - S^m_{a'b'} \rangle' \quad . \tag{2.6}$$

It is noted that, according to eq(2.6), all the coupling coefficients in eq (2.4) are invariant with respect to inversion of the solvent molecular frame $(x', y', z') \leftrightarrow (-x', -y', -z')$ and would therefore survive for an achiral nematic solvent as well as a chiral one but would not change on inverting the handedness of a chiral solvent. Consequently, the inversion-symmetric, second rank contribution of eq(2.4) to the potential of men torque describes the part of the solute ordering that is independent of the solvent chirality, in short, the achiral part of the ordering.

To construct the lowest rank contributions that would vanish for an achiral nematic solvent but not for a chiral one and would change on changing the sense of chirality of the solvent, we consider the vector-pseudovector contributions of the form

$$V^{(p)}(\omega) = \bar{C}^{(p)}_{a \times b, c; a' \times b', c'} \left( \left\langle \left( (\hat{a} \times \hat{b}) \cdot \hat{c}' \right) \left( (\hat{a}' \times \hat{b}') \cdot \hat{c} \right) \right\rangle' - \frac{1}{3} \varepsilon_{abc} \varepsilon_{a'b'c'} \right) \quad , \quad (2.7)$$

With the Levi-Civita symbol $\varepsilon_{abc} = \pm 1$ for cyclic/anticyclic permutations of $x, y, z$ and vanishing otherwise. Taking into account the $D_2$ phase symmetry, the complete vector-pseudovector contribution of eq (2.7) can be written in the form:

$$V^{(p)}(\omega) = \left\langle g^{(p)}_{a \times b, c} \right\rangle' A^n_{a \times b, c} + \left\langle \Delta^{(p)}_{a \times b, c} \right\rangle' \left( A^l_{a \times b, c} - A^m_{a \times b, c} \right) \quad , \quad (2.8)$$

where, in analogy with the definitions in eqs (2.6), (2.7) we have for the orientational vector-pseudovector components and the respective coupling coefficients, uniaxial and biaxial, the definitions:

$$\left\langle g^{(p)}_{a \times b, c} \right\rangle' \equiv \bar{C}^{(p)}_{a \times b, c; a' \times b', c'} \left\langle A^n_{a' \times b', c'} \right\rangle' \; ; \; \left\langle \Delta^{(p)}_{a \times b, c} \right\rangle' \equiv \frac{1}{3} \bar{C}^{(p)}_{a \times b, c; a' \times b', c'} \left\langle A^l_{a' \times b', c'} - A^m_{a' \times b', c'} \right\rangle' \quad . \quad (2.9)$$

Each of the above coupling coefficients changes sign under the reversal of the solvent molecular axes $(x', y', z') \leftrightarrow (-x', -y', -z')$. Therefore, the $V^{(p)}(\omega)$ contribution to the potential of mean torque given in eq (2.8), to be termed henceforth as the inversion-antisymmetric contribution, vanishes for achiral solvents and reverses its sign on changing the handedness of chiral solvents.

As demonstrated in Sectionς 4 and 5, using for the potential of mean torque just the "achiral" contributions given in eq (2.4) and the "chiral" contribution of eq (2.8), i.e. taking

$$V(\omega) = V^{(2)}(\omega) + V^{(p)}(\omega) \quad , \quad (2.10)$$

leads to the correct description of the available results of enantiotopic NMR discrimination for the different possible molecular symmetries of achiral solutes and provides very precise reproductions of the experimental measurements on rigid solutes.

## 3. NMR SPECTRA AND MOLECULAR ORIENTATIONAL ORDERING

In this work we shall consider rigid solute molecules studied by NMR techniques based on (a) proton –proton/carbon dipolar couplings and (b) deuterium quadrupole couplings with the bond electric field gradient. In both cases, the NMR spectral signature $v_s$ of a molecular segment $s$ (bond vector or inter-nuclear vector) of direction $\hat{e}_s$ is generally

$$v_s = v_s^0 \left\langle P_2(\hat{H} \cdot \hat{e}_s) \right\rangle \quad , \tag{3.1}$$

where $v_s^0$ is a site-specific constant and $\hat{H}$ denotes the direction of the magnetic field. $P_2(x) = (3x^2 - 1)/2$ denotes the second Legendre polynomial and the orientational average indicated by the angular brackets is obtained from the angular distribution $f(\omega)$ of the solute molecule in the ordered medium as

$$\left\langle X \right\rangle \equiv \int X(\omega) f(\omega) d\omega \ . \tag{3.2}$$

Specifically [1]:

(a) The dipolar coupling between a pair of nuclei is given as:

$$D_{ij} = -K_{i-j} \left\langle \frac{1}{r_{ij}^3} P_2(\hat{H} \cdot \hat{r}_{ij}) \right\rangle \tag{3.3}$$

where $r_{ij}$ denotes the distance between the $i$-$j$ pair of nuclei, $\hat{r}_{ij}$ the direction of the internuclear vector, $K_{i-j} \equiv \mu_o h \gamma_i \gamma_j / 16\pi^3$ is the dipolar coupling constant for the $i$, $j$ pair with gyromagnetic ratios $\gamma_i, \gamma_j$ and $\mu_0$ is the vacuum permeability. Thus, for dipolar couplings of strictly rigid molecules, the spectral signature $v_s$ corresponds to $D_{ij}$ and therefore $v_s^0 = -K_{i-j} / r_{ij}^3$ and the site direction $\hat{e}_s$ corresponds to the internuclear unit vector $\hat{r}_{ij}$. We have $K_{C-H} = 30.0$ kHzÅ$^3$ and $K_{H-H} = 120.9$ kHzÅ$^3$ for the C-H and H-H pairs, respectively.

(b) The deuterium nucleus interacts, through its electric quadrupolar moment, with the electric field gradient (EFG) across the nucleus [1]. Typically, for a deuterium bonded to an aliphatic carbon atom the EFG tensor is essentially axially symmetric about the direction of the C-D bond and the spectroscopically observable quadrupolar splitting associated with such bond is given by

$$\Delta v_{CD} = \frac{3}{2} q_{CD} \left\langle P_2(\hat{H} \cdot \hat{e}_{CD}) \right\rangle \tag{3.4}$$

where $q_{CD}$ ($\approx 167 kHz$ for aliphatic carbons) is the quadrupolar coupling constant and the unit vector $\hat{e}_{CD}$ is in the direction of the C-D bond. In this case, the spectral signature $\nu_s$ corresponds to $\Delta \nu_{CD}$; therefore $\nu_s^0 = 3 q_{CD}/2$ and the site direction $\hat{e}_s$ corresponds to the bond unit vector $\hat{e}_{CD}$. The EFG tensors are slightly asymmetric for the aromatic carbons and the respective quadrupolar coupling constants for aromatic carbons ($q_{CD} \approx 185\ kHz$) are typically 10% higher than those of aliphatic deuterons.

Provided that the solute molecules are completely rigid, the generalized spectral signature of eq (3.1) can be expressed in terms of purely geometric molecular constants, describing the orientation of the site vector $\hat{e}_s$ in the solute molecular frame $x, y, z$, and of certain orientational averages of the magnetic field direction relative to $x, y, z$, as follows:

$$\begin{aligned}
\nu_s / \nu_s^0 &= \langle P_2(\hat{H} \cdot \hat{e}_s) \rangle \\
&= P_2(\hat{z} \cdot \hat{e}_s) \langle P_2(\hat{H} \cdot \hat{z}) \rangle + \frac{3}{4}\left((\hat{x} \cdot \hat{e}_s)^2 - (\hat{y} \cdot \hat{e}_s)^2\right)\left\langle (\hat{x} \cdot \hat{H})^2 - (\hat{y} \cdot \hat{H})^2 \right\rangle + \\
&\quad 3 \begin{pmatrix} (\hat{x} \cdot \hat{e}_s)(\hat{y} \cdot \hat{e}_s) \langle (\hat{x} \cdot \hat{H})(\hat{y} \cdot \hat{H}) \rangle + (\hat{x} \cdot \hat{e}_s)(\hat{z} \cdot \hat{e}_s) \langle (\hat{x} \cdot \hat{H})(\hat{z} \cdot \hat{H}) \rangle + \\ (\hat{z} \cdot \hat{e}_s)(\hat{y} \cdot \hat{e}_s) \langle (\hat{z} \cdot \hat{H})(\hat{y} \cdot \hat{H}) \rangle \end{pmatrix}
\end{aligned} \quad (3.5)$$

Generally, the direction of the spectrometer magnetic field $\hat{H}$ relative to the director frame $\hat{n}, \hat{l}, \hat{m}$ may vary within the solvent phase. However, in the experimental cases of interest to this work, the primary director $\hat{n}$ of the phase is uniformly oriented parallel to the applied magnetic field. For a common uniaxial nematic (N) medium this would correspond to a magnetically aligned positive diamagnetic anisotropy sample while for a chiral nematic (N*) it would correspond to the unwinding of the cholesteric helix under the action of the magnetic field. In all such cases, the direction of the magnetic field in eq (3.5) can be identified with the director $\hat{n}$. Therefore, the expression of the NMR signatures can be written in terms of the second rank order parameters of the solute molecule, $\langle S_{ab}^n \rangle$, and the respective site geometrical constants

$$S_{ab}^s = \frac{3}{2}(\hat{a} \cdot \hat{e}_s)(\hat{b} \cdot \hat{e}_s) - \frac{1}{2}\delta_{ab} \quad , \quad (3.6)$$

as follows

$$\nu_s / \nu_s^0 = S_{zz}^s \langle S_{zz}^n \rangle + \frac{1}{3}\left(S_{xx}^s - S_{yy}^s\right)\left(\langle S_{xx}^n \rangle - \langle S_{yy}^n \rangle\right) + \frac{4}{3}\left(S_{xy}^s \langle S_{xy}^n \rangle + S_{xz}^s \langle S_{xz}^n \rangle + S_{yz}^s \langle S_{yz}^n \rangle\right) \quad (3.7)$$

A pair of sites $s, s'$ which are related by a symmetry operation of the molecule will obviously have identical spectral constants $\nu_s^0 = \nu_{s'}^0$. If, furthermore, this symmetry operation is a proper rotation, bringing the direction $\hat{e}_s$ onto $\hat{e}_{s'}$ and vice versa, then the two sites will exhibit identical spectral signatures, $\nu_s = \nu_{s'}$. This follows directly from eq (3.7) and the potential of mean torque in eq (2.11) upon identifying the $z$ molecular axis with the symmetry axis, leading automatically to $\langle S_{xz}^n \rangle = \langle S_{yz}^n \rangle = 0$ and $S_{zz}^{s'} = S_{zz}^s$. Also, the arbitrariness in the directions of the $x, y$ axes can be lifted by requiring that $\langle S_{xy}^n \rangle = 0$. Thus, of the solute order parameters appearing in eq (3.7), we are left with only $\langle S_{zz}^n \rangle$ and $\langle S_{xx}^n \rangle - \langle S_{yy}^n \rangle$. The latter order parameter would also vanish if the symmetry axis $z$ is higher than twofold and could generally assume non vanishing values if $z$ is a twofold symmetry axis but then the geometrical constants $S_{xx}^{s(s')} - S_{yy}^{s(s')}$ would be identical for the two sites $s, s'$; in either case, therefore, the spectral signatures of the two sites would be equal.

### 4. ENANTIOTOPIC DIFFERENTIATION AND SOLUTE MOLECULAR SYMMETRY

Having demonstrated directly that there can be no spectral discrimination between molecular sites which are related by proper symmetry operations, we now turn to the spectral signatures of molecular sites which are related by improper symmetry operations (reflections and improper rotations), i.e. enantiotopic sites. These are considered in detail for all the possible molecular symmetries that are compatible with the assumed achirality of the solute molecules. The symmetry elements of such molecules necessarily include an axis of improper n-fold rotation $S_n$. Accordingly, here we scan through the possible point groups of molecular symmetry containing such improper rotations and consider their implications on the differentiation between the spectral signatures of two sites $s, s'$ related by improper symmetry operations.

**4.1 Mirror plane** ($C_s = S_1$). Identifying the $z$ molecular axis with the symmetry plane normal (therefore the $x, y$ axes are on the symmetry plane) we have that three of the five independent geometrical constants in eq (3.7) are identical for the two mirror-image sites $s, s'$ and the remaining two geometrical constants are of opposite sign, namely,

$$S_{xz}^{s'} = -S_{xz}^s \; ; \; S_{yz}^{s'} = -S_{yz}^s \tag{4.1}$$

For the same identification of the $z$ axis, and the $x, y$ axes on the symmetry plane but otherwise arbitrary, the nonvanishing terms in the symmetric part of the potential of mean torque, $V^{(2)}(\omega)$ in eq (2.4), are those with the index pairs $ab = xx / yy / zz / xy$, for the antisymmetric part, $V^{(p)}(\omega)$ in eq (2.8), the surviving terms are those with the index triplets $a \times b, c = x \times y, x / x \times y, y$. Accordingly, the total potential of mean torque $V(\omega)$ in eq (2.10), leads to nonvansihing values of all the independent solute order parameters $\langle S^n_{ab} \rangle$ in eq (3.7) and therefore, taking into account eq (4.1) the spectral difference between the $s, s'$ in this case

$$\delta_{ss'} \equiv (\nu_s - \nu_{s'})/\nu_s^0 = \frac{8}{3}\left(S^s_{xz}\langle S^n_{xz}\rangle + S^s_{yz}\langle S^n_{yz}\rangle\right) \tag{4.2}$$

**4.2 Mirror plane and symmetry axis normal to it ($C_{nh}$).** It is apparent from the previous Section that the presence a n-fold symmetry axis, with $n \geq 2$ along the $z$ molecular axis implies the vanishing of $S^{s'}_{xz}, S^{s'}_{yz}, S^s_{xz}, S^s_{yz}$ as well as of the remaining terms of the $V^{(p)}(\omega)$ part and thereby of both order parameters $\langle S^n_{xz}\rangle, \langle S^n_{yz}\rangle$. Accordingly there is no NMR differentiation (i.e. $\delta_{ss'} = 0$) between enantiotopic sites in the case of $C_{nh}$ symmetry with any $n \geq 2$. Obviously, this holds also for point croups containing additional symmetry elements ($D_{nh}$).

**4.3 Mirror planes containing a symmetry axis ($C_{nv}$).** Identifying the $z$ molecular axis with the $C_n$ axis, and, without loss of generality, the $x$ axis to be on a mirror plane we have, for a pair of sites $s, s'$ which are enatiotopic with respect to that plane,

$$S^{s'}_{xz} = -S^s_{xz} \; ; \; S^{s'}_{xy} = -S^s_{xy}, \tag{4.3}$$

with the remaining three geometrical constants in eq (3.7) being identical for the two sites. The nonvanishing terms in the symmetric part of the potential of mean torque, are those with the index pairs $ab = xx / yy / zz$. For the antisymmetric part, the surviving terms for $n = 2$ are those with the index triplets $a \times b, c = x \times z, x / y \times z, y$, implying the nonvanishing of the solute order parameter $\langle S^n_{xy}\rangle$ while for $n > 2$, all the terms vanish, leading to $\langle S^n_{xy}\rangle = 0$. Therefore, for $C_{nv}$ molecular symmetry, there is differentiation between enantiotopic sites only for n=2, and the quantitative expression is given by

$$\delta_{ss'} = 8S^s_{xy}\langle S^n_{xy}\rangle/3 \quad , \tag{4.4}$$

with the $x,y$ axes taken normal to the two mirror planes.

**4.4 Extension to $D_{nd}$.** This can be done by adding to the symmetry elements of the $C_{nv}$ point group $n$ twofold axes bisecting the dihedral angles of the mirror planes. Obviously, the absence of spectral differentiation shown in the previous Section for $n > 2$ will remain. For $n = 2$, the axtenssion to $D_{2d}$ by adding two $C_2$ axes, which according to the axis assignment in Section 4.3 bisect the angles formed by the $x$ and $y$ axes, imply the invariance under the simultaneous transformation $(x, y, z) \leftrightarrow (\pm y, \pm x, -z)$. This is in addition to the invariance implied by the $C_{2v}$ symmetry, i.e. under independent inversions $x \leftrightarrow -x$ and $y \leftrightarrow -y$. Accordingly, for the two enantiotopic sites of Section 4.3, the relations between their orientation constants remain the same (see eq (4.3)). The nonvanishing terms in the symmetric part of the potential of mean torque remain the same as in Section 4.3, i.e. $ab = xx/yy/zz$, with the added symmetry constraint that the contributions of the $xx$ and $yy$ terms are equivalent, (i.e. $\langle g^{(2)}_{xx} \rangle' = \langle g^{(2)}_{yy} \rangle'$ ; $\langle \Delta^{(2)}_{xx} \rangle' = \langle \Delta^{(2)}_{yy} \rangle'$). Accordingly, the symmetric part of the potential of mean torque has effectively only a $zz$ contribution for $D_{2d}$. The antisymmetric part also has the same nonvanishing contributions as in Section 4.3, but again with equivalent contributions for the terms differing by the exchange of the $x,y$ axes, namely $\langle g^{(p)}_{x \times z,x} \rangle' = -\langle g^{(p)}_{y \times z,y} \rangle' = \langle g^{(p)}_{z \times y,y} \rangle'$ ; $\langle \Delta^{(p)}_{x \times z,x} \rangle' = \langle \Delta^{(p)}_{z \times y,y} \rangle'$. Altogether then, there is differentiation` between the two sites for n=2 and the quantitative expression for $\delta_{ss'}$ is the same as in eq (4.4) inly in this case the order parameters are evaluated with the $D_{2d}$-symmetric potential of mean torque which is invariant with respect to the exchange of the $x$, $x,y$ axes.

4.5. **Roto-reflections ($S_{2n}$).** Lastly we consider the combined *2n*-fold rotation and reflection about a mirror lane normal to the rotation axis. The $S_2$ symmetry is the same as the inversion symmetry which makes all the terms of the antisymmetric part $V^{(p)}(\omega)$ vanish, leading to the vanishing of $\delta_{ss'}$. Similarly, the $S_{2n}$ symmetry for $n>2$ leads vanishing of $V^{(p)}(\omega)$ and therefore the only case of interest for enantiotopic spectral differentiation is $S_4$. With $z$ identified again with the $C_4$ axis, this symmetry implies invariance under the simultaneous transformations of the molecular axes $(x, y, z) \leftrightarrow (\pm y, \mp x, -z)$. This leads to a $V^{(2)}(\omega)$ with only a $zz$ contribution and a $V^{(p)}(\omega)$ in which the only nonvanishing terms are $\langle g^{(p)}_{x \times z,y} \rangle' = \langle g^{(p)}_{y \times z,x} \rangle'$ ; $\langle g^{(p)}_{x \times z,x} \rangle' = -\langle g^{(p)}_{y \times z,y} \rangle'$ and $\langle \Delta^{(p)}_{x \times z,y} \rangle' = \langle \Delta^{(p)}_{y \times z,x} \rangle'$ ; $\langle \Delta^{(p)}_{x \times z,x} \rangle' = -\langle \Delta^{(p)}_{y \times z,y} \rangle'$. Thus, $V^{(2)}(\omega)$ alone would give rise to only $\langle S^I_{zz} \rangle \neq 0$ order parameters for the solute molecule while the presence of $V^{(p)}(\omega)$ provides the additional order parameters $\langle S^I_{xx} - S^I_{yy} \rangle \neq 0$ and $\langle S^I_{xy} \rangle \neq 0$. Furthermore, two sites related by a fourfold roto-reflection, $S_4$, symmetry operation have identical geometrical constants $S^{s'}_{zz} = S^s_{zz}$ and opposite signs for $S^{s'}_{xx} - S^{s'}_{yy} = -\left(S^s_{xx} - S^{e_s}_{yy}\right)$ ; $S^{s'}_{xy} = -S^s_{xy}$. Thus the two sites are spectrally differentiated and the quantitative expression is in the case of $S_4$ solute molecular symmetry

$$\delta_{ss'} = \frac{2}{3}\left(S^s_{xx} - S^s_{yy}\right)\left(\langle S^n_{xx} \rangle - \langle S^n_{yy} \rangle\right) + \frac{8}{3} S^s_{xy} \langle S^n_{xy} \rangle.$$

In summary, the enantiotopic discrimination can be observed on the NMR spectra only for solutes of molecular symmetry corresponding to one of the four point groups: $C_s$, $C_{2v}$, $D_{2d}$, and $S_4$. This is in agreement with all the available experimental observations [8] and with conclusions reached through a different line of reasoning, based on the quenching of the effective molecular symmetry of a rigid solute molecule when dissolved in a chiral medium.

## 5. CALCULATION OF ENANTIOTOPIC SPECTRAL DIFFERENTIATION AND COMPARISON WITH MEASUREMENT

**5.1 Acenaphthene.** To illustrate the use of the potential of mean torque in a chiral nematic solvent, formulated in Section 2, we consider first its application to the evaluation of the qudrupolar couplings of the rigid molecule acenaphthene, the ordering of which, particularly in chiral nematic media formed by PBLG/CHCl$_3$, has been widely studied by deuterium NMR spectroscopy [8], [14]. The molecular geometry of acenaphthene is shown in figure 2. According to eq (3.4), the deuterium spectral frequencies provide quantitative experimental evaluation of the segmental order parameters associated with the C-D bond molecular directions. Taking into account that phase biaxiality in the usual chiral nematic solvents (N$^*$) is rather small, we shall ignore in our calculations the biaxial contributions to both, the symmetric $V^{(2)}(\omega)$ and the antisymmetric $V^{(p)}(\omega)$ parts of the potential of mean torque.

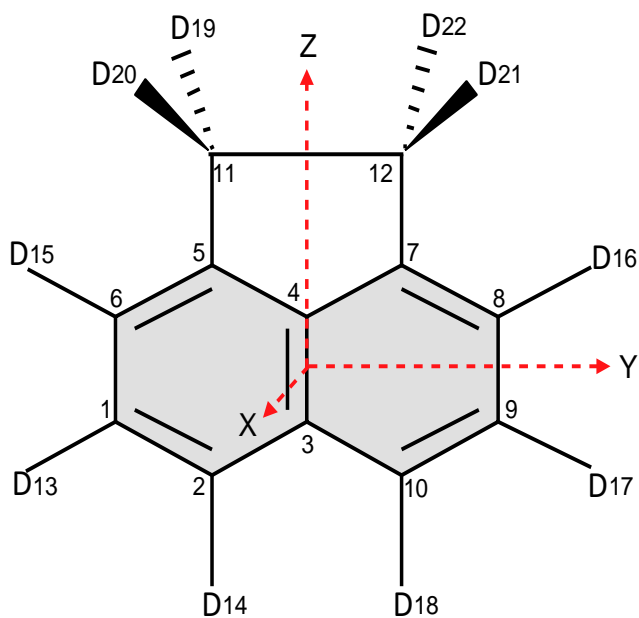

**Figure 2**: Molecular structure of acenaphthene showing the site labeling and the molecular frame of axes used in the calculation of residual dipolar couplings and deuterium quadrupolar couplings.

The acenaphthene molecule belongs to the $C_{2v}$ point group symmetry. It therefore follows from Section 4.3 that the potential of mean torque, restricted to uniaxial contributions, is formed by the symmetric part

$$V^{(2)}(\omega) = \left(\langle g_{zz}^{(2)}\rangle' - \frac{1}{2}\left(\langle g_{xx}^{(2)}\rangle' + \langle g_{yy}^{(2)}\rangle'\right)\right)S_{zz}^n + \frac{1}{2}\left(\langle g_{xx}^{(2)}\rangle' - \langle g_{yy}^{(2)}\rangle'\right)\left(S_{xx}^n - S_{yy}^n\right) \quad, \tag{5.1}$$

and the antisymmetric part

$$V^{(p)}(\omega) = \langle g_{x\times z,x}^{(p)}\rangle' A_{x\times z,x}^n + \langle g_{y\times z,y}^{(p)}\rangle' A_{y\times z,y}^n \quad. \tag{5.2}$$

With $\theta$ and $\varphi$ denoting the polar and azimuthal angles of the direction of the director $\hat{n}$ relative to the right-handed molecular frame $x$, $y$, $z$, the total potential of mean torque can be but in the explicitly angle-dependent form

$$V(\omega) = V^{(2)}(\omega) + V^{(p)}(\omega) = u_0^{(2)}\left(\frac{3}{2}\cos^2\theta - \frac{1}{2}\right) + u_2^{(2)}\sin^2\theta\cos 2\varphi + u_{12}^{(p)}\sin^2\theta\sin 2\varphi \quad, \tag{5.3}$$

where the solvent-order-dependent parameters are given by:

$$u_0^{(2)} = \langle g_{zz}^{(2)}\rangle' - \frac{1}{2}\left(\langle g_{xx}^{(2)}\rangle' + \langle g_{yy}^{(2)}\rangle'\right) \;;\; u_2^{(2)} = \frac{3}{4}\left(\langle g_{xx}^{(2)}\rangle' - \langle g_{yy}^{(2)}\rangle'\right) \;;\; u_{12}^{(p)} = \frac{3}{4}\left(\langle g_{y\times z,y}^{(p)}\rangle' - \langle g_{x\times z,x}^{(p)}\rangle'\right) \tag{5.4}$$

It is noted that (i) according to the inversion symmetry properties of the coupling coefficients in eqs (2.6) and (2.10), inverting the chirality of the solvent has no effect on $u_0^{(2)}, u_2^{(2)}$ but inverts the sign of $u_{12}^{(p)}$ and (ii) the term $u_{12}^{(p)}\sin^2\theta\sin 2\varphi$ leads to different values of the probability distribution for mirror image directions of $\hat{n}$ relative to the symmetry planes of the molecule. This, of course, does not imply any change of the molecular symmetry; it merely conveys the symmetry of the molecular ordering in a medium that, being chiral, distinguishes between opposite faces of a mirror plane. Equivalently, the $\varphi$ dependence of the potential of mean torque in eq (5.4) can be put in the form $u_2^{(2)}\sin^2\theta\cos 2\varphi + u_{12}^{(p)}\sin^2\theta\sin 2\varphi = \sqrt{\left(u_2^{(2)}\right)^2 + \left(u_{12}^{(p)}\right)^2}\sin^2\theta\cos 2(\varphi - \varphi_0)$, and made to appear as a potential which is symmetric in a frame sharing the same $z$ axis with the actual (i.e. symmetry defined) molecular but having the other two axes rotated by an angle $\varphi_0 = \frac{1}{2}\tan^{-1}\left(u_{12}^{(p)}/u_2^{(2)}\right)$ relative to the actual $x,y$ axes.

With the numbering used in Figure 2, the enantiotopic deuterium pairs are {$D_{19}$, $D_{20}$} and {$D_{21}$, $D_{22}$} while the two pairs are related by a twofold rotation about the $z$ axis and therefore the spectra of $D_{19}$ and $D_{21}$ are identical and so are those of $D_{20}$ and $D_{22}$. The relevant geometrical constants can be evaluated directly from the known geometry of the molecule [14]; these are $S_{zz}^{C-D_{20}} = S_{zz}^{C-D_{19}} = -0.166$, $S_{xx}^{C-D_{20}} - S_{yy}^{C-D_{20}} = S_{xx}^{C-D_{19}} - S_{yy}^{C-D_{19}} = 0.782$, $S_{xy}^{C-D_{20}} = -S_{xy}^{C-D_{19}} = -0.154$, and identically for $D_{21}$, $D_{22}$. The quadrupolar frequency splittings for any one of these deuterated sites $s=19,20,21,22$ of acenaphthene are given according to eqs (3.4) and (3.7) by

$$\Delta \nu_{19} = (3q_{CD}/2)\left[ S_{zz}^s \langle S_{zz}^n \rangle + \frac{1}{3}(S_{xx}^s - S_{yy}^s)(\langle S_{xx}^n \rangle - \langle S_{yy}^n \rangle) + \frac{4}{3} S_{xy}^s \langle S_{xy}^n \rangle \right] \quad (5.5)$$

Table 1 shows the results of the calculation of the frequency splittings in the quadrupolar spectra associated with the five inequivalent deuterated sites of the acenaphthene molecule at various temperatures in the chiral nematic solvent PBLG/CHCl$_3$. The calculated values are optimized relative to the experimentally measured values by varying $u_0^{(2)}$, $u_2^{(2)}$, $u_{12}^{(p)}$ in the potential of mean torque of eq (5.3); these are the only fitting parameters in the calculation. Their so obtained optimal values are shown in the graphs of Figure 3. According to these optimal values, the ordering matrix of the molecule becomes diagonal when the principal (i.e. symmetry defined) molecular axis frame $x,y,z$ of figure 2 is rotated about the $z$-axis by an angle $\varphi_0$ in the range $1.6°$ to $1.9°$, with a slight tendency to increase with increasing temperature.

**Table 1**. Comparison between experimental [14] data (in the parenthesis) and calculated results of quadrupolar splittings for acenaphthene in the chiral nematic solvent PBLG/CHCl$_3$ within the temperature range 295-330°K.

| T/K | $\Delta\nu_{19}$ or $\Delta\nu_{20}$ | $\Delta\nu_{20}$ or $\Delta\nu_{19}$ | $\Delta\nu_{16}= \Delta\nu_{15}$ | $\Delta\nu_{17}= \Delta\nu_{13}$ | $\Delta\nu_{18}= \Delta\nu_{14}$ |
|---|---|---|---|---|---|
| 295 | -376.5(-377.0) | -419.4(-420.0) | 484.3(485.0) | 483.6(482.0) | 494.5(494.0) |
| 300 | -359.6(-360.0) | -401.5(-402.0) | 460.4(461.0) | 459.2(458.0) | 480.4(480.0) |
| 305 | -344.8(-345.0) | -384.7(-385.0) | 438.1(439.0) | 436.2(435.0) | 469.3(469.0) |
| 310 | -332.7(-333.0) | -370.7(-371.0) | 419.9(421.0) | 417.5(416.0) | 459.3(459.0) |
| 315 | -317.9(-318.0) | -354.9(-355.0) | 399.5(401.0) | 396.6(395.0) | 445.2(445.0) |
| 320 | -306.0(-306.0) | -342.1(-342.0) | 383.0(385.0) | 379.8(378.0) | 434.1(434.0) |
| 325 | -293.2(-293.0) | -329.2(-329.0) | 365.6(367.0) | 362.0(361.0) | 422.9(423.0) |
| 330 | -280.3(280.0) | -315.4(-315.0) | 348.1(350.0) | 344.2(343.0) | 409.8(410.0) |

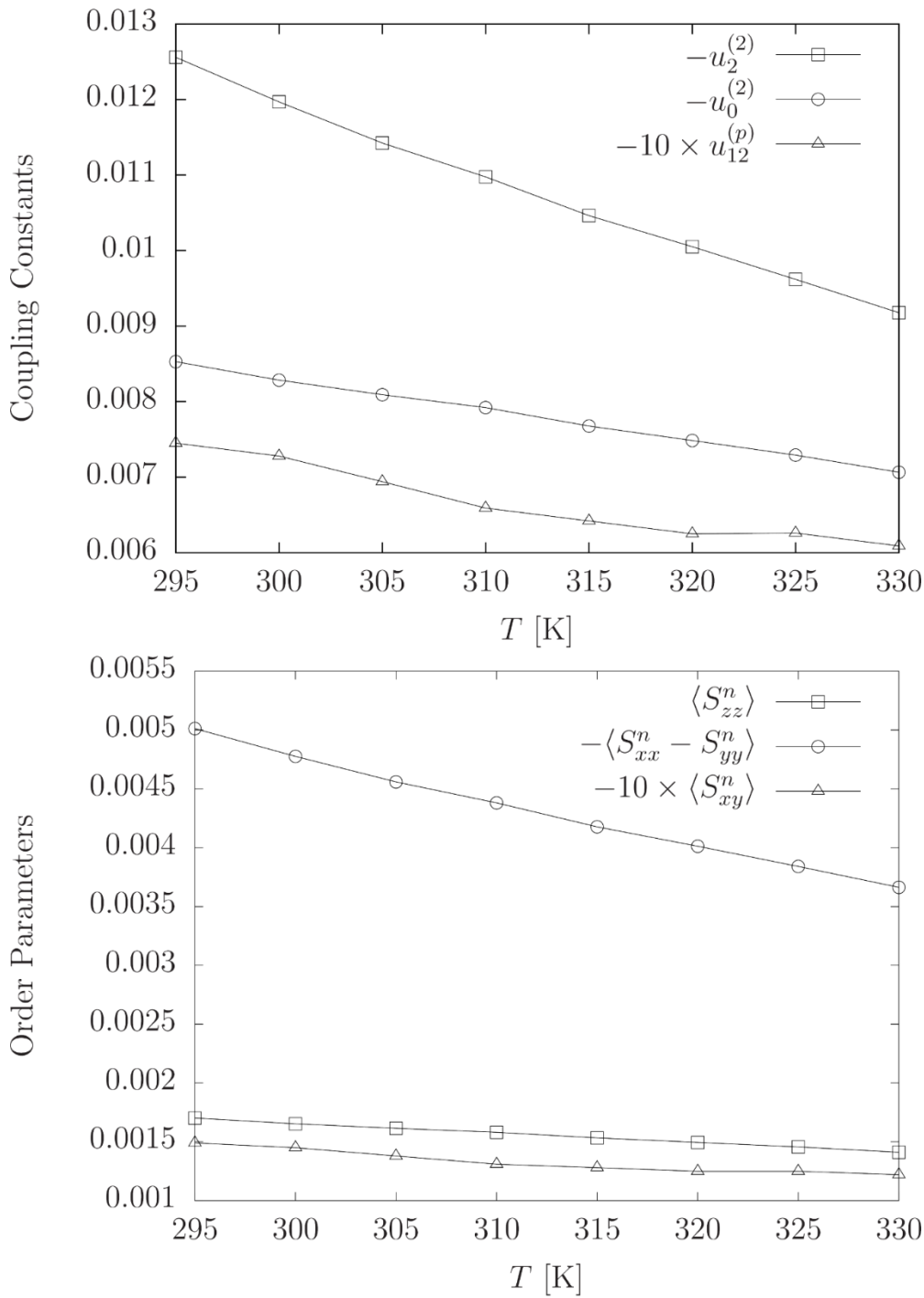

**Figure 3.** Temperature dependence of the optimal values of the three coupling parameters of the potential of mean torque of eq (5.4) for acenaphthene dissolved in PBLG/CHCl$_3$ (Top) and of the corresponding order paramenteres entering the expression for the quadrupolar splittings in eq (5.5)(bottom).

It is apparent from Table 1 that the agreement with experiment is excellent for all the measured splittings. The temperature dependence of the fitting parameters in Figure 3 is in accord with their relation to the ordering of the solvent molecules, shown in eqs (2.6), (2.10), and (5.4). Also, the calculated solute order parameters in the same figure follow the expected temperature dependence, their relative magnitudes are consistent with the choice of the molecular *x* axis normal to the planar core of the molecule.

**5.2 Ethanol.** As shown in Section 4, the lowest molecular symmetry for the observation of enentiotopic differentiation is the presence of just a mirror plane. According to Section 4.1, the potential of mean torque for a rigid molecule of $C_s$ symmetry, neglecting the biaxial contributions, would consist of the terms

$$V(\omega) = \left(\left\langle g_{zz}^{(2)}\right\rangle' - \frac{1}{2}\left(\left\langle g_{xx}^{(2)}\right\rangle' + \left\langle g_{yy}^{(2)}\right\rangle'\right)\right)S_{zz}^n + \frac{1}{2}\left(\left\langle g_{xx}^{(2)}\right\rangle' - \left\langle g_{yy}^{(2)}\right\rangle'\right)\left(S_{xx}^n - S_{yy}^n\right)$$
$$+ \left\langle g_{xy}^{(2)}\right\rangle' S_{xy}^n + \left\langle g_{x\times y,x}^{(p)}\right\rangle' A_{x\times y,x}^n + \left\langle g_{x\times y,y}^{(p)}\right\rangle' A_{x\times y,y}^n \qquad (5.6)$$

For a right handed frame of orthogonal molecular axes *xyz*, this can be put in a form analogous to eq (5.3), namely

$$V(\omega) = u_0^{(2)}\left(\frac{3}{2}\cos^2\theta - \frac{1}{2}\right) + u_2^{(2)}\sin^2\theta\cos 2\varphi + u_1^{(2)}\sin^2\theta\sin 2\varphi$$
$$+ \left(u_{13}^{(p)}\cos\varphi + u_{23}^{(p)}\sin\varphi\right)\sin 2\theta \qquad (5.7)$$

where $u_0^{(2)}, u_2^{(2)}$ are defined in eq (5.4), $u_1^{(2)} = 3\left\langle g_{xy}^{(2)}\right\rangle'/4$ and $u_{13}^{(p)} = 3\left\langle g_{x\times y,x}^{(p)}\right\rangle'/4$, $u_{23}^{(p)} = 3\left\langle g_{x\times y,y}^{(p)}\right\rangle'/4$. Thus, the potential of mean torque, for this low symmetry case, is specified in terms of minimally five parameters. To determine these by fitting experimental spectra in the manner described in the previous Section for the acenaphthene molecule, would require a sufficiently restrictive set of data. However, to our knowledge, no set of NMR data providing five or more independent spectral quantities is available in the literature for strictly rigid solute molecules of $C_s$ symmetry. On the other hand, ethanol, although not a completely rigid molecule, has been studied extensively by NMR [8], [15] and a sufficiently restrictive set of independent dipolar couplings and quadrupolar splittings have been measured in the solvents PBLG/CHCl$_3$ [15] and PCBLL/CHCl$_3$ [16]. It is therefore of some interest to examine whether the use of the potential of mean torque that is applicable to rigid molecules would provide a reasonable description for molecules which are neither strictly rigid nor highly flexible. This is done here for ethanol.

The molecular structures of ethanol in its *trans* conformation is shown in Figure 4. Ethanol is a not a rigid molecule due to two internal motions: one is the 3-fold symmetric rotation of the methyl group about the C-C bond and the other is the rotation of the hydroxyl group about the C-O bond. The methyl group rotation does not influence the ordering properties of the molecule in any significant way and its effects on the relevant bond directions and interatomic distances can be readily taken into account. The situation is different with the rotations about the C-O bond and for that reason a systematic scanning has been performed to obtain the respective torsional energy profiles. There are three minima of the torsion energy, corresponding to the *trans* conformation, shown in Figure 4, for which the torsion angle is defined at $\alpha_t^{CO} = 180°$ and the two mirror image *gauche* conformations at about $\alpha_{g\pm}^{CO} = \pm 63°$. The energy difference between the *trans* and *gauche* conformations ($E_{tg}$) has been calculated to be 0.60 Kcal/mol. These results are in accord with those obtained by other ab initio calculations [17], [18]. The structural parameters for the optimized geometry of ethanol, including bond lengths, bond angles, and dihedral angles were obtained by using the slandered DFT/B3LYP Method [19]. The bond length of the central C-C bond was found to be 1.514Å while the C-H bonds of methyl and methylene groups were found to be 1.095Å. The angle between two methylene C-H directions (H7-C2-H8) was found to be 109.1°.

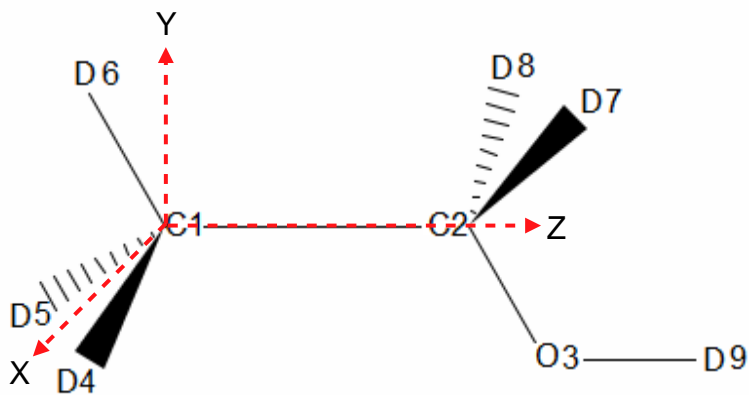

**Figure 4**: Molecular structure of ethanol with atomic labeling and the molecular axis frame used in the calculations.

From the analysis of proton, C-13, and deuterium NMR spectra, D Merlet et al [15] reported the chiral discrimination in ethanol dissolved in PBLG/CHCl$_3$. Here, we have employed the potential of mean torque given in eqs (5.6) and (5.7) to reproduce the experimental data of the dipolar couplings measured in the $^1$H and $^{13}$C NMR spectra in [15]. It should be noted that the use of this form of $V(\omega)$ implies that the three distinct conformations of ethanol are assumed to have identical orientational distribution functions and furthermore that the molecule has a plane of symmetry. The latter is strictly true only for the *trans* state of the molecule and could be assumed to be valid in a statistical sense when the two *gauche* states occur with equal probability. However, the *gauche* states are chiral enantiomers and would therefore exhibit different ordering tendencies in a chiral environment and different overall probabilities between $g^+$ and $g^-$. Also, the orientational constants $S_{ab}^s$ can be readily obtained from the molecular geometry only for those segments *s* which do not change on changing conformation; for the segments whose geometry does depend on the molecular conformations, $S_{ab}^s$ cannot be treated as constants and their values are replaced by averages over the conformations of the isolated molecule, i.e. ignoring the aforementioned deviations which are caused by the chiral environment.

The fitting was performed by varying the five parameters of the potential of mean torque in eq (5.6) to minimize the overall scaled deviation of the calculated values from the measured ones, $\sum \left( \frac{D_{ij}^{(\exp)} - D_{ij}^{(calc)}}{\sigma_{ij}^{(\exp)}} \right)^2$, with the summation running over all the pairs for which experimental data are available and $\sigma_{ij}^{(\exp)}$ denoting the reported experimental uncertainty for the dipolar coupling of the ij pair. A comparison of experimental [15] and calculated dipolar couplings is presented in Table 2 for the ethanol dissolved in PBLG/CHCl₃. The respective optimal values of the fitting parameters are listed in Table 3.

**Table 2.** Experimental [15] and calculated dipolar couplings for ethanol (in Hz) in PBLG/CHCl$_3$ at temperature 300K.

| Atom Types | Interacting Nuclei | $D_{ij}$(**exp**) | $D_{ij}$(**cal**) |
|---|---|---|---|
| C-H | 1-7 | 3.40(0.60) | 3.43 |
| C-H | 1-8 | 3.40(0.60) | 3.02 |
| C-H | 1-(4,5,6) | -4.59(0.10) | -4.51 |
| C-H | 2-7 | 24.94(0.10) | 24.93 |
| C-H | 2-8 | 10.18(0.10) | 10.17 |
| C-H | 2-(4,5,6) | 1.12(0.10) | 1.22 |
| H-H | 7-8 | 23.65(0.10) | 23.66 |
| H-H | 7-(4,5,6) | 4.76(0.50) | 4.78 |
| H-H | 8-(4,5,6) | 4.76(0.50) | 4.79 |
| H-H | 4,5,6 | -6.78(0.10) | -6.76 |

**Table 3:** Optimal values of the potential of mean torque parameters obtained from fitting the experimental dipolar couplings of ethanol [15] in PBLG/CHCl$_3$ shown in table 2.

| $u_0^{(2)}$ | $u_2^{(2)}$ | $u_1^{(2)}$ | $u_{13}^{(p)}$ | $u_{23}^{(p)}$ |
|---|---|---|---|---|
| 0.00318 | 0.00713 | -0.01209 | -0.00456 | -0.00026 |

The calculated results are in excellent agreement with the experimental data of dipolar couplings. In fact, the experimental data are fully reproduced within the experimental accuracy. As seen in Table 2, the experimental dipolar couplings for the two enantiotopes of the methylene group, $D_{2-7}$ and $D_{2-8}$, are different. Our calculated values of $D_{2-7}$ and $D_{2-8}$ are also found to be distinct and they are in good agreement with the experimental values. Furthermore, two other possibilities of enantiotopic pairs exist in ethanol: one is 1-7 and 1-8 and the other is (7-4,5,6) and (8-4,5,6). Experimentally, no differences were obtained between the values of dipolar couplings for $D_{1-7}$ and $D_{1-8}$ nor for the $D_{7-(4,5,6)}$ and $D_{8-(4,5,6)}$ pairs. This result has been interpreted to imply that the differences in couplings, if present, are too small to be observed experimentally [15]. By using instead of PBLG a different chiral orienting medium, PCBLL, Aroulanda et al [16] were able to differentiate, in terms of dipolar couplings, the two enantiotopic directions 7-(4,5,6) and 8-(4,5,6) but not the 1-7 and 1-8 directions. However our calculations depict a small difference between the enantiotopic directions 7-(4,5,6) and 8-(4,5,6) as well as for the 1-7 and 1-8 directions.

Next, we calculated the quadrupolar splittings, using directly the optimal values of the parameters in Table 3 and without any further optimization of the molecular structure (for example, to take into account the replacement of the protons with deuterons). The results are presented in Table 4 and are compared with the experimentally measured values for the ethanol dissolved in PBLG/CHCl$_3$ [15].

**Table 4** Experimental [15] and theoretical quadrupolar splittings for ethanol (in Hz) in PBLG/CHCl$_3$ at temperature 300K.

| Atom Types | Interacting Nuclei | $\Delta v_Q$ (**exp**) (Hz) | $\Delta v_Q$ (**cal**) (Hz) |
|---|---|---|---|
| C-D | 1-(4,5,6) | 53.3 | 50.0 |
| C-D | 2-8 | -136.1(a) | -114.1 |
| C-D | 2-7 | -264.2(b) | -280.5 |
| O-D | 3-9 | 765.8 | |

Whilst the magnitudes and the experimental trends are reproduced correctly, there are appreciable deviations of our calculated values from the experimental data for the C-D bonds. Clearly, these deviations go beyond possible fine tuning regarding, for example, the direction of the principal axis of the EFG tensor not coinciding exactly with the C-D bond (our calculations indicate a small deviation of ~1-2°) or some spread (of up to 20kHz) in the values of the quadrupolar coupling constant $q_{CD}$ reported in the literature. Moreover, the literature values of the quadrupolar coupling constant $q_{OD}$ for an O-D bond present a rather broad spread [20], extending over the range of 50-300 kHz, with reported values up to 328 kHz [21] and possibly influenced by hydrogen bonding. Due to the high uncertainty for the quadrupolar coupling constant of the O-D bond, its asymmetry and the orientation of the respective EFG tensor principal axes, a definite value of quadrupolar splitting could not be calculated reliably. However, the calculated uniaxial order parameter of the O-D is not influenced by such uncertainty and is found to be $\langle P_2(\hat{n} \cdot \hat{e}_{O-D}) \rangle \approx 7.5 \times 10^{-4}$.

Summarizing the use of the rigid $C_s$-symmetric potential of mean torque of eq (5.6) for the description of ethanol, we note that while the parameters of the potential can be optimized to provide an essentially exact reproduction of the dipolar couplings, the so obtained optimal values yield a reasonable but not as accurate reproduction of the quadrupolar splittings. This lack of perfect transferability of the optimal parameters from one set of measurements to the other is attributed primarily to the neglect of the small differences in the ordering tendencies and overall probabilities of the conformational states generated by the rotations of the O-D bond. A systematic treatment of flexible solute molecules by means of a conformation dependent potential of mean torque is presented in a subsequent publication [part II: flexible solutes]

## 6. CONCLUSIONS

The potential of mean torque, and thereby the orientational distribution function, for rigid solute molecules dissolved in chiral nematic solvents has been formulated in terms molecular coupling parameters with clear connection to the low-rank solute-solvent tensorial interactions, the symmetry of both types of molecules and the macroscopic symmetry of the solvent phase. The inclusion of inversion-antisymmetric ("chiral") terms is shown to reproduce correctly all the symmetry point groups which lead to NMR-detectable enantiotopic discrimination for rigid prochiral solute molecules in chiral nematic solvents. These terms are also shown not only to reproduce rather accurately the measured spectra for the strictly rigid solute acenaphthene in PBLG/$CHCl_3$ chiral environments but also to provide a fairly accurate description of the spectra for ethanol, with remarkable resolution for enantiotopic pairs of nuclei, despite the lack of complete rigidity of the molecular structure. The molecular symmetry of the solute molecules is explicit and transparent at all stages of the formulation, up to its final application for the reproduction of the experimental data. The difference between the solute orientational distribution, and thereby of the order parameters, in chiral and achiral nematic media follows clearly and explicitly from the tensorial character of the terms contained in the potential of mean torque and no quenching of the solute molecular symmetry is involved at any point. The approach we have presented here for rigid solutes is extended in a forthcoming publication to flexible molecules dissolved in chiral nematic solvents.

**ACKNOWLEDGMENT**

This research has been co-financed by the European Union (European Social Fund – ESF) and Greek national funds through the Operational Program 'Education and Lifelong Learning' of the National Strategic Reference Framework (NSRF) – Research Funding Program: THALES [grant number MIS: 380170].